\documentclass[aps,prb,superscriptaddress,twocolumn,longbibliography
]{revtex4-2}

\usepackage{amsmath, amsthm, amssymb}
\usepackage{graphicx}
\usepackage{times}
\usepackage{xcolor}
\usepackage{xargs}
\usepackage[colorlinks,citecolor=blue,urlcolor=blue]{hyperref}
\usepackage[utf8]{inputenc}
\usepackage{ragged2e}
\usepackage[colorlinks,citecolor=blue,urlcolor=blue]{hyperref}

\begin{document}

\title{On-demand higher-harmonic generation through nonlinear Hall effects in curved nanomembranes}

\author{Botsz Huang}
\affiliation{Department of Physics, National Cheng Kung University, Taiwan}
\affiliation{Center for Quantum Frontiers of Research and Technology (QFort), National Cheng Kung University, Tainan 70101, Taiwan}

\author{You-Ting Huang}
\affiliation{Department of Physics, National Cheng Kung University, Taiwan}
\affiliation{Center for Quantum Frontiers of Research and Technology (QFort), National Cheng Kung University, Tainan 70101, Taiwan}

\author{Jan-Chi Yang}
\affiliation{Department of Physics, National Cheng Kung University, Taiwan}
\affiliation{Center for Quantum Frontiers of Research and Technology (QFort), National Cheng Kung University, Tainan 70101, Taiwan}

\author{Tse-Ming Chen}
\affiliation{Department of Physics, National Cheng Kung University, Taiwan}
\affiliation{Center for Quantum Frontiers of Research and Technology (QFort), National Cheng Kung University, Tainan 70101, Taiwan}

\author{\\Ali~G.~Moghaddam}\email{a.ghorbanzade@gmail.com}
\affiliation{Department of Physics, Institute for Advanced Studies in Basic Sciences (IASBS), Zanjan 45137-66731, Iran}
\affiliation{Computational Physics Laboratory, Physics Unit, Faculty of Engineering and
Natural Sciences, Tampere University, FI-33014 Tampere, Finland}

\author{Ching-Hao Chang}\email{cutygo@phys.ncku.edu.tw}
\affiliation{Department of Physics, National Cheng Kung University, Taiwan}
\affiliation{Center for Quantum Frontiers of Research and Technology (QFort), National Cheng Kung University, Tainan 70101, Taiwan}

\date{\today} 

\begin{abstract}
The high-order Hall effects, which go beyond the ordinary, unlock more possibilities of electronic transport
properties and functionalities. Pioneer works focus on the manufacture of complex nanostructures with low
lattice symmetry to produce them. In this paper, we theoretically show that such high-order Hall effects can
alternatively be generated by curving a conducting nanomembrane which is highly tunable and also enables
anisotropy. Its Hall response can be tuned from first to fourth order by simply varying the direction and magnitude
of the applied magnetic field. The dominant Hall current frequency can also be altered from zero to double,
or even four times that of the applied alternating electric field. This phenomenon is critically dependent on
the occurrence of high-order snake orbits associated with the effective magnetic-field dipoles and quadruples
induced by the curved geometry. Our results offer pathways for spatially engineering magnetotransport, current
rectification, and frequency multiplication in the bent conducting nanomembrane.
\end{abstract}

\maketitle

\section{Introduction}
The Hall effect, which was first observed over a century and a half ago as a transverse voltage in a conductor subjected to a perpendicular magnetic field, has become a cornerstone in modern solid-state physics \cite{Hall}. The discovery of integer and fractional quantum Hall phenomena, along with the later predictions and observations of spin Hall effects, have expanded the importance of the Hall effect at both fundamental and practical levels. It has a rich underlying physics, connecting to topology, quantum information, and the physics of strong correlations, and provides concrete applications, from magnetic field sensing to precise measurements of fundamental constants, as well as potential uses in nanoelectronics, spintronics, and quantum computation.

Recently, a new aspect has been added to the field with the prediction and experimental confirmation of the nonlinear Hall effect in systems that break inversion symmetry while preserving time-reversal symmetry \cite{PhysRevB.99.155404,PhysRevLett.123.246602,carmine2021review,du2021perspective,Ma_2018,Kang_2019,PhysRevLett.123.036806,shvetsov2019nonlinear,huang2020giant}.
The Berry curvature dipole, a unifying key concept, governs the existence and strength of the second-order Hall effect. There has been a recent endeavor to find materials with a large Berry curvature dipole
\cite{PhysRevB.97.035158, PhysRevB.97.041101, PhysRevB.98.121109, Du2018, PhysRevLett.121.246403, PhysRevLett.123.196403, PhysRevLett.124.067203, PhysRevLett.125.046402, PhysRevB.102.245116, polini2018, Juridic2020, wawrzik2020infinite, malla2021emerging}. 
Moreover, the high-harmonics components of Hall effect are explored to be driven in the strained 2D materials recently \cite{Ho_2021,Kong_2022,Tamaya_2023}.
Its generation is associated with nonlinear time-dependent and AC transport and optical phenomena and thus has a high potential in application
\cite{isobe2020high,PhysRevApplied.13.024053,PhysRevB.102.245422,Kumar2021,he2021quantum,gao2021second,zhang2021terahertz,Kruchinin_2018,Park_2022,Borsch_2023}.

In addition to the nonlinear Hall effect driven by the Berry curvature dipole in specific materials,  it is noteworthy that the ordinary Hall effect with higher order can also be externally induced by an applied magnetic field dipole \cite{Huang2021}. Our study centers on the nonlinear ordinary Hall effect due to its distinctive application potentials, as both its occurrence and order can be externally manipulated in conventional materials \cite{Ramsden2006, Goel2020, Klitzing2020, Huang2021}.

\begin{figure*}[tb]
\centering 
\includegraphics[width=.8\textwidth]{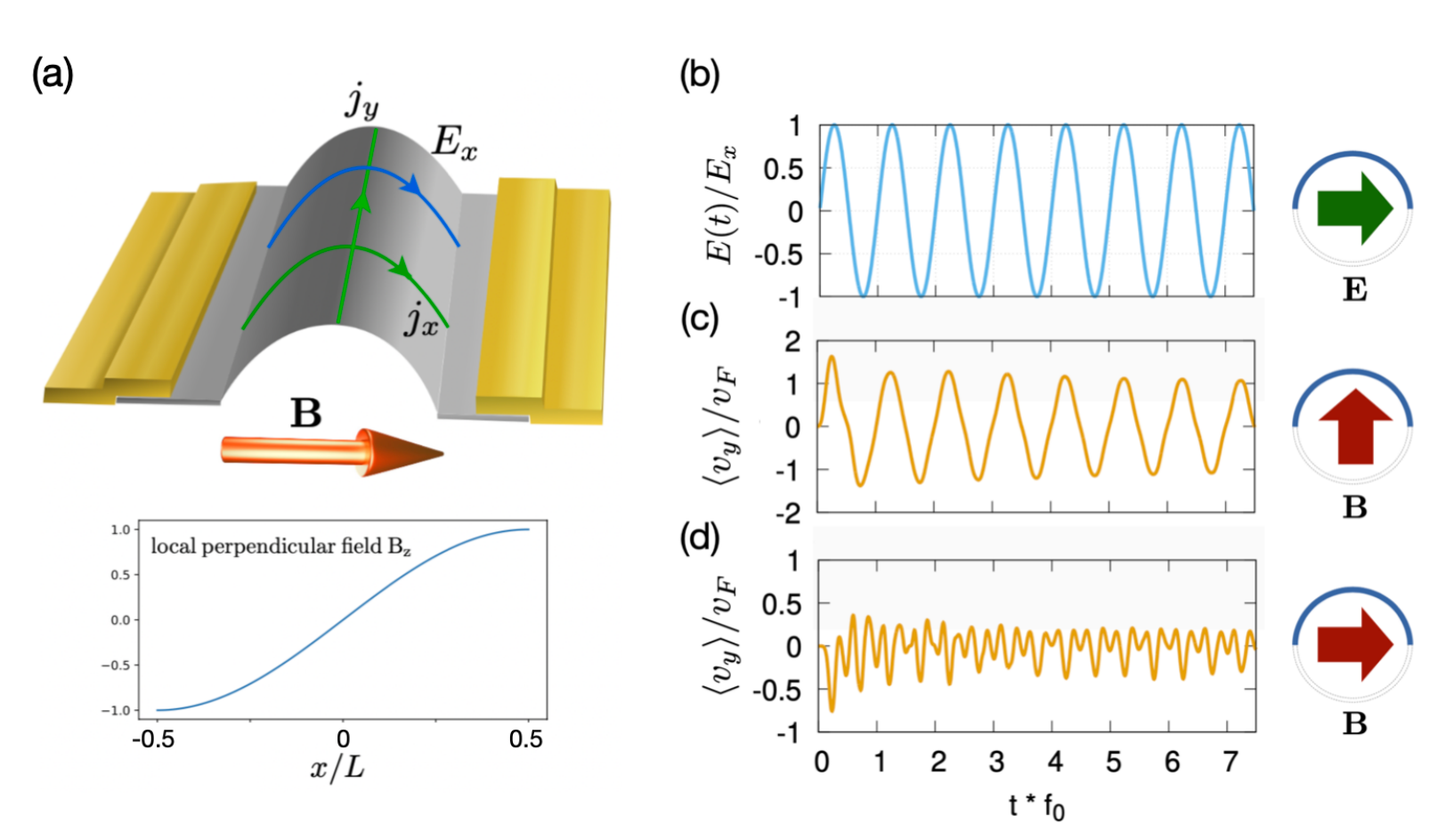}
\caption{(Color online) 
Switching between ordinary and nonlinear Hall effect can be easily achieved by rotating the applied magnetic field ${\bf B}$. (a) shows a schematic of the proposed device. By applying an AC electric field with a frequency $f_0$ as shown in (b)
we also get a transverse signal which can be dominated by ordinary linear Hall effect or nonlinear behavior depending on the magnetic field orientation as shown in (c) and (d), respectively. When ${\bf B}$ is along the $z$ direction, the 
transverse signal almost follows the applied electric field with the same frequency (c), whereas for in-plane alignment as ${\bf B} = B_0\hat{\bf x}$,
the electronic gas in semi-cylindrical region experiences an effective $B$-dipole
and as a result we have a nonlinear response where higher harmonics dominate the transverse signal.
Here $x$, $y$ and $z$ are the axes of Cartesian coordinates on the curved nanomembrane.
}
\label{fig1}
\end{figure*}

In the paper, the Hall effect is examined in two-dimensional electron gases (2DEGs) when the system is curved as schematically illustrated in Fig. \ref{fig1}(a). By applying a uniform constant magnetic field to this curved system, the effective perpendicular magnetic field which controls Hall response of the system will be inhomogeneous due to the curvature.
We find that by varying only the orientation and also strength of the applied magnetic field, we can get rich behavior for the Hall effect, particularly we can tune between different nonlinear orders and particularly \emph{second}- to \emph{forth}- harmonic components. The main underlying reason for this behavior is that by changing the orientation of applied field we can induce a real-space magnetic field dipole or even quadruple which subsequently drives a nonlinear Hall effect.
Exploiting a semiclassical framework and numerical simulation
based on that, we can trace the spatiotemporal evolution and variation of the Hall current, which demonstrate the competition of \emph{zeroth}-, \emph{second}-, and \emph{forth}- harmonic components in the device for in-plane orientation of magnetic field. The high sensitivity and controllability of the high-order Hall effect and the order of dominant harmonics make this device very promising for exploring nonlinear Hall effect beyond the more commonly studied second-order case. 

\section{Semiclassical approach}\label{}
Our theoretical framework to study the higher-order Hall effects in curved nanomembranes is based on  semiclassical equations of motion. 
The role of curvature is taken into account through the position-dependence of magnetic field. Hence, we need to numerically evaluate the solution for the semiclassical equations
  \begin{equation}\label{eq:m1}
       \dot{\mathbf r} = \frac{\partial 
       \mathcal{E}(\mathbf p)}{\partial {\mathbf p}} 
       \end{equation}
    \begin{equation}\label{eq:m2}
      \dot{\mathbf p} = - \frac{e}{\hbar} \left[ \mathbf{E (t)}- { \dot {\mathbf r}} \times { \mathbf B ( \mathbf{r}) }  \right]
    \end{equation}
for a 2DEG with band dispersion $\mathcal{E}$ in the presence of a generic position-dependent magnetic field ${\bf B}({\bf r})$.
Here $\mathbf r$ and $\mathbf p$ represent position and momentum of the center of an electron wave-packet, respectively,
and $\mathbf{E}$ indicates the applied electric field which drives the electrons to move.
The methodology of the simulation is provided in Sec. S1 in the Supplemental Material.
Assuming a homogeneous 2DEG model near the band edge we essentially deal with a single band with quadratic energy dispersion where we can also omit the band indices.
To obtain the equilibrium, the equations of motion are supplemented with a Boltzmann approach, as explained in Refs. \cite{mahan2013many, PhysRevB.59.14915, PhysRevLett.112.166601}. Following the semiclassical approach, our theoretical study is implemented by a test particle method to simulating the impact of magnetic field on electron dynamics numerically.

We consider the clean limit, $\Gamma \rightarrow 0$, and the Hall current is computed by numerically integrating the trajectories of independent particles governed by Eqs.\eqref{eq:m1} and \eqref{eq:m2}. 
This clean limit requires that the mean free path of material is longer than the arc length $L = 1 \mu m$ of the curved nanomembrane and thus impact of impurities and defects in transport is negligible. Ballistic transport regime with vanishingly small scattering rate ($\Gamma$) is achievable in modern two-dimensional conducting materials \cite{beenakker1991, Banszerus_2016}.
This numerical self-evolved approach samples the distribution function in the clean limit and enables the computation of the full harmonics of the Hall current, without the requirement of being close to the weakly nonlinear regime \cite{testparticles}.
A series of time-dependent numerical simulations are performed with an electric field ${\bf E}(t) =\hat{\bf x}\: E_x(t)= \hat{\bf x}\: E \sin (\omega t)$, with the frequency $f_0 = \omega / 2\pi $ in the THz range. 
Here two million test particles are employed which are uniformly placed in the $x$-axis in the spatial space and with a Fermi velocity of $v_F = 10^6 $ m/s in the beginning of simulation.
The elastic collisions are taken into account for test particles on the boundary in real space.
By means of statistical analysis, the Hall response is determined by taking the average of the momentum $p_y$.

\section{Results}\label{}
The schematic of our proposed device is shown in Fig. \ref{fig1}(a), where we have a curved 2D layer of electron gas subjected to an external magnetic field whose direction can be varied. Assuming a semicylindrical shape for the curved region, the magnetic field perpendicular to the electron gas will change according to $B_{z}(x)= B_0\sin (\pi x/L+\theta)$, where $\theta$ indicates the angle between the magnetic field direction and the $x$ axis. The resulting magnetic field for $\theta=0$ is also shown in the inset of Figs. \ref{fig1}(a). Using the numerical method explained earlier, we calculated the current passing in longitudinal and vertical directions as a result of an AC electric field in the presence of effective, spatially-varying magnetic field, as shown in Fig. \ref{fig1}(c) and (d). The effective $B_z$ field plays an important role in the Hall effect in the curved nanomembrane. The simulation in Fig. \ref{fig1}(c) shows the ordinary Hall effect when the effective $B_z$ field is always in the same direction. 
Consequently, the ordinary linear Hall effect dominates the transport  
in this situation. By rotating the magnetic field, such that its perpendicular component changes sign as shown in the inset of Fig. \ref{fig1}(d), there is no net perpendicular magnetic field on average. Instead, we have a real-space magnetic field dipole that can give rise to a nonlinear Hall effect and higher-order harmonic components, as shown in Ref. \cite{Huang2021}. 

\begin{figure}[tb!]
\includegraphics[width=.47\textwidth]{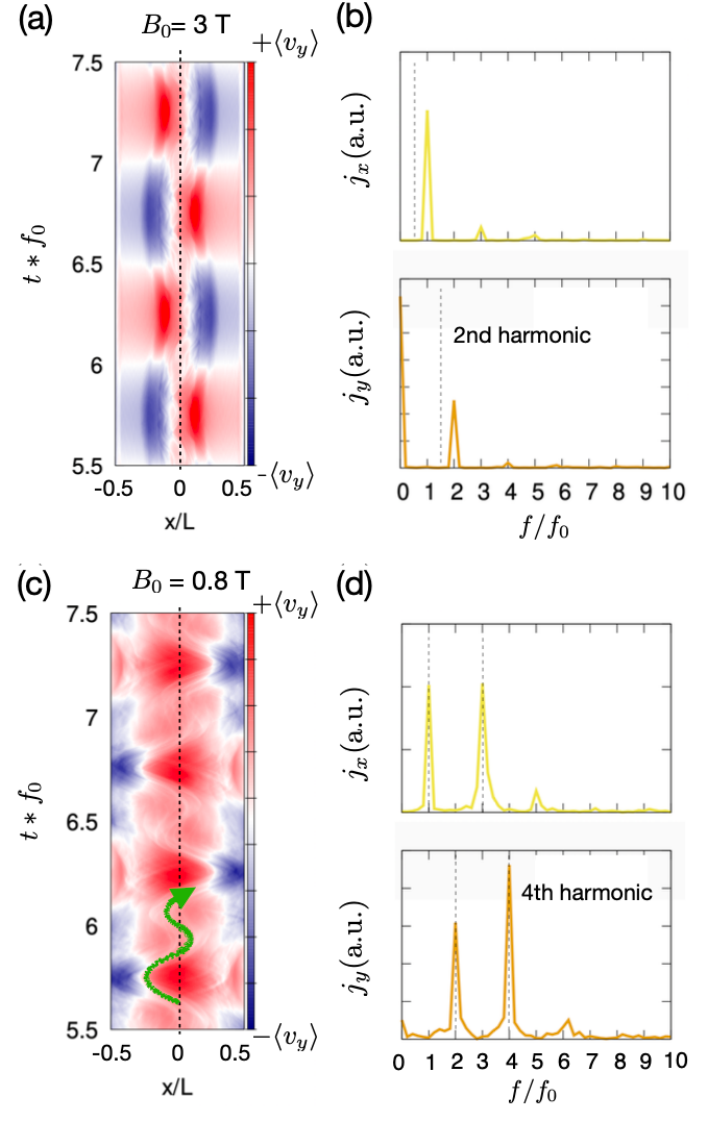}
\caption{(Color online) The frequencies of longitudinal and Hall current depend on the strength of applied magnetic field. (a) Simulation of the distribution and velocity of magnetic orbits in the nanomembrane as the function of time for the strong $B$-field dipole. The dashed line indicates the position of $x/L=0$ where $B_z = 0$. (b) Comparing with the frequency of AC voltage, the frequencies of longitudinal current $j_x$ is the same, and the major frequencies of Hall current $j_y$ are zero and two times.
(c) Simulation of the distribution and velocity of magnetic orbits as the function of time for the moderate field. The solid green line shows the snake orbits. (d) Comparing with the frequency of AC voltage, the major frequencies of longitudinal current $j_x$ are one and three times, and the major frequencies of Hall current $j_y$ are two and four times.}
\label{fig2}
\end{figure}

Our semiclassical simulations can also easily be used to show the spatiotemporal 
distribution profile of the transverse velocity which further justifies the existence of higher harmonics as can be seen in Fig. \ref{fig2}.
This results show how the curved nanomembrane supports higher harmonics of current in both longitudinal and transverse directions, when the magnetic field is applied along the $x$ direction.
In addition to the second-harmonic response dominating transverse Hall currents when the applied field $B_0 > 3$ T [see the lower panel in Fig. \ref{fig2}(b)], we find, in particular, a very strong fourth-harmonic component of transverse Hall currents when the field $B_0 \approx 0.8$ T, as shown in the lower panel in Fig. \ref{fig2}(d).
Accordingly, odd-order harmonic components appear in the longitudinal current, with the first and third harmonics being the dominant contributions [see the upper panel in Fig. \ref{fig2}(d)].
While a second harmonic Hall response has been observed in various systems and situations, higher components have not typically been found to be strong in many models and previous studies. In our research, we show that by choosing a suitable curved electronic device and tuning magnetic field strength, we can generate very strong fourth-harmonic transverse current through the spatial variations of the effective perpendicular magnetic field.

\begin{figure}[tb!]
  \centering 
\includegraphics[width=.47\textwidth]{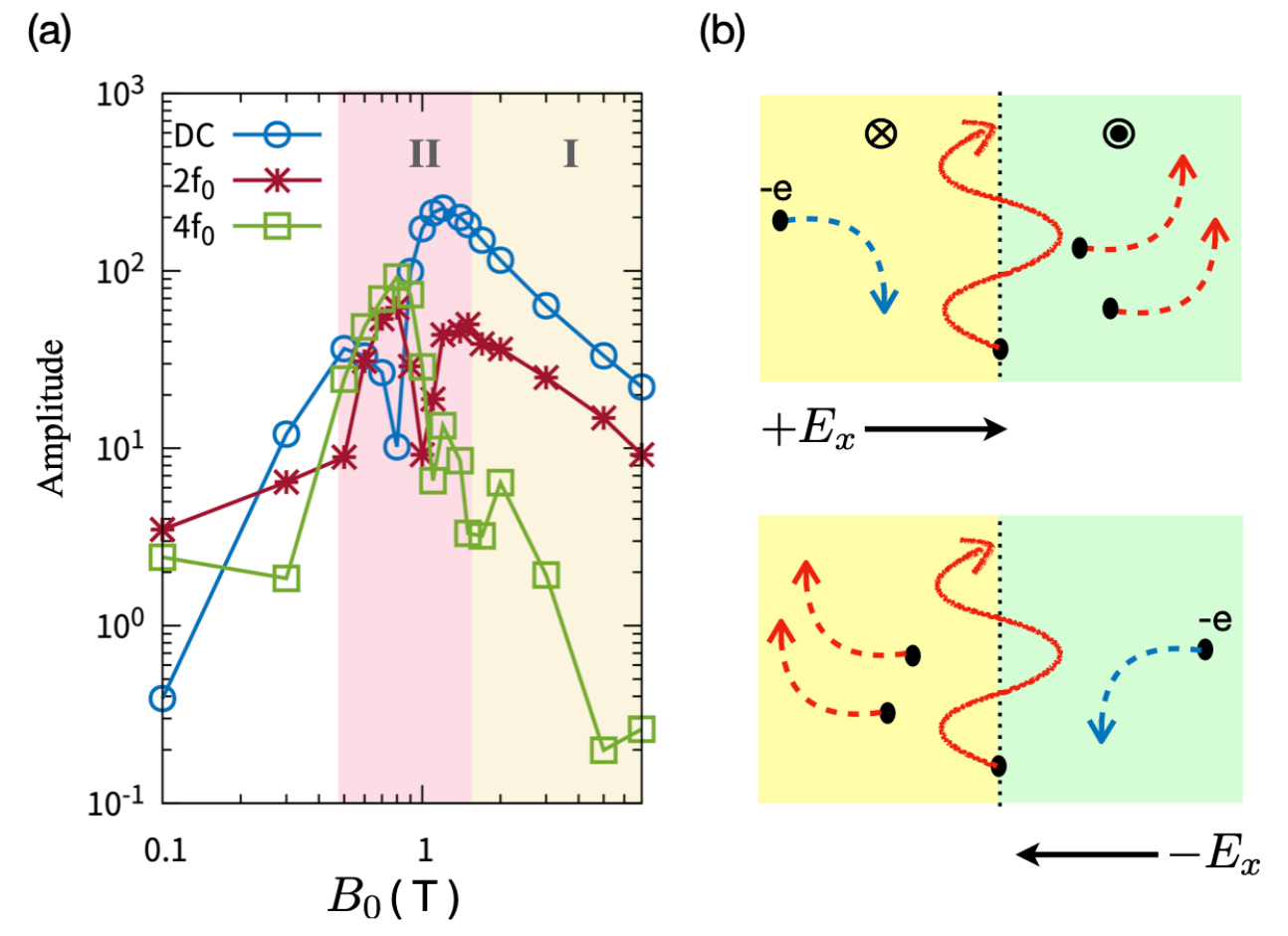}
\caption{(Color online) (a) The high order Hall effects due to the $B$-field dipole, following the nature of the ordinary Hall effect, drive the current with zero, double and four times of frequencies. (b) The formation of drift orbits (snake orbits) in the $B$-field dipole contributes to the Hall current with a frequency that is two (four) times of that of AC electric field. Finally, the Hall current flows along the $+y$ direction whether the applied field is along the $+x$ (upper panel) or $-x$ (lower panel) directions. }
\label{fig3}
\end{figure}

By varying the magnitude of the magnetic field, 
the amplitude of different nonlinear terms corresponding to higher harmonic generation, also significantly changes. This is elucidated in Fig. \ref{fig3}(a) for the relative strength of the zeroth- (DC term), second-, and fourth-harmonic Hall signals as functions of applied field strength.
In this figure, we observe a resonant-like feature associated with the strength of the Hall response as a function of the magnetic field. We can see two prototypical sections labeled as I and II. When the $B$ field is strong (I), the DC term and then second-harmonic components dominate the Hall current, while higher harmonics, especially the fourth-harmonic contribution, are significantly suppressed. Note that we are using a logarithmic scale in Fig. \ref{fig3}(a). Near the resonant-like behavior labeled with II, the fourth harmonic of the Hall response increases dramatically, and it can even be the largest contribution for a range of magnetic fields. This result shows a major finding of our work: we can engineer a dominant fourth harmonic generation in the Hall response of a curved electronic system by simply tuning the magnitude of the applied magnetic field.

To understand the origin of nonlinear Hall response and the appearance of strong second and fourth harmonics, we consider two different regimes of large and intermediate magnetic fields, separately.
In the limit of strong magnetic field, labeled as part I in Fig. \ref{fig3}(a), the snake orbits illustrated in Fig. \ref{fig3}(b) are localized at the vicinity of $x/L = 0$. 
We note that snake orbit is the high-order magnetic trajectory for a charge carrier moving across the interface where magnetic field changes its sign (magnetic field dipole) \cite{Mueller_1992,Reijniers_2000}. The contribution of these snake orbits in the magnetotransport can be neglected in this limit and therefore the magnetic profile around this region ($x/L \sim 0$) can be well approximated with a dipole\footnote{This mainly originates from the fact that expanding $\sin(x)$ function around $x=0$ results in a linear dependence. 
Therefore in this limit magnetic profile in the region of interest is $B(x) \sim  x$ representing a real-space magnetic dipole.}.
Thus, the Hall current is almost generated by the difference of carrier density distribution between the areas with effective $+B_z$ and $-B_z$ field [see Figs. \ref{fig1}(a) and \ref{fig3}(b))], and can be approximated as \cite{MacDonald1984},
\begin{align}\label{eq:3}
\notag j_y &\cong 
j_y^{(1)}\big|_{R}  \frac{\mu_R}{\mu} - j_y^{(1)}\big|_{L}  \frac{\mu_L}{\mu}\\
& \propto e^2  \frac{E^2_x(t)}{B_0}
\end{align}
where $j_y^{(1)}$ is the ordinary Hall current for carrier density $n$ driven by the local magnetic field. Here $\mu_{R}$ and $\mu_{L}$ is chemical potential at the right and left side, respectively, and $\mu = (\mu_R + \mu_L) / 2$ is the averaged chemical potential.
From the above equation we derived (Sec. S2 in the Supplemental Material), we can readily see the origin of nonlinear Hall effect in this system with a sign-changing magnetic field profile whereby the net Hall current of the full system is at least proportional to the square of electric field. 

This explanation is indeed consistent with the numerical results shown in Fig. \ref{fig3}(a) for region I, where we have strong magnetic field and the magnetic profile is dominated only by the dipole contribution. In this limit the main contribution to the Hall current is the second-order term, given with the expression $j_y^{(2)} \propto {E}^2_x(t)/B_0$, which can be decomposed into two Fourier components corresponding to the DC term and the second-harmonic term. The amplitude of second-harmonic term inversely proportional to the field strength $B_0$ is also shown in our numerical results (see Fig. S2(b) in the Supplemental Material).

So far, we have established that the 2$^{\rm nd}$-order Hall currents depicted in Fig. \ref{fig2}(b) arise from the interplay between the inverse components of ordinary Hall effects in a nanomembrane and an applied magnetic-field dipole, as shown in Fig. \ref{fig2}(a). This behavior relies on the magnetic field being strong enough to disregard the influence of higher-order magnetic orbits, such as snake orbits that are confined to a narrow region at the interface of the $B$-field dipole, as depicted in Fig. \ref{fig2}(a) and Fig. \ref{fig3}(b). However, when the applied magnetic-field strength is moderate, snake orbits significantly impact magnetotransport because they are dispersed across the nanomembrane, as illustrated in Fig. \ref{fig2}(c). 
In this regime, the semiclassical trajectory of particles becomes more involved and snake orbits further spread in perpendicular direction such that electrons are influenced by the higher magnetic moments. This follows the simple fact that by going further away from the middle of the nanomembrane where the effective magnetic 
field changes its sign, the deviation of magnetic profile from a linear variation becomes more apparent.
In particular, the contribution of quadruple magnetic moment in the Hall dynamics of particles increase which yields a strong fourth-harmonic component.

The dynamic of snake orbits is intrinsically different with drift orbits for they only moves along the a transverse direction, as illustrated in Fig. \ref{fig3}(b). The transverse velocity is $v_y \propto  \sin^2(\omega t)$ because they only move along $+y$ and oscillate coherently with ${\bf E}(t)$ [see Fig. \ref{fig2}(c)]. As an extension of second-order Hall currents in Eq.(\ref{eq:3}) with a revision of transverse velocity $v_y$, fourth-harmonic component occurs majorly when $B_0 \approx 0.8$ T, as is shown in region II in Fig. \ref{fig3}(a). The key nature of the amplitude of fourth-harmonic component, namely $j_y^{(4)} \propto {E}_x^4$, is identified in our numerical results (see Sec. S3 in the Supplemental Material).

\section{Discussion}\label{}
\begin{figure}[tb!]
  \centering 
\includegraphics[width=1\columnwidth]{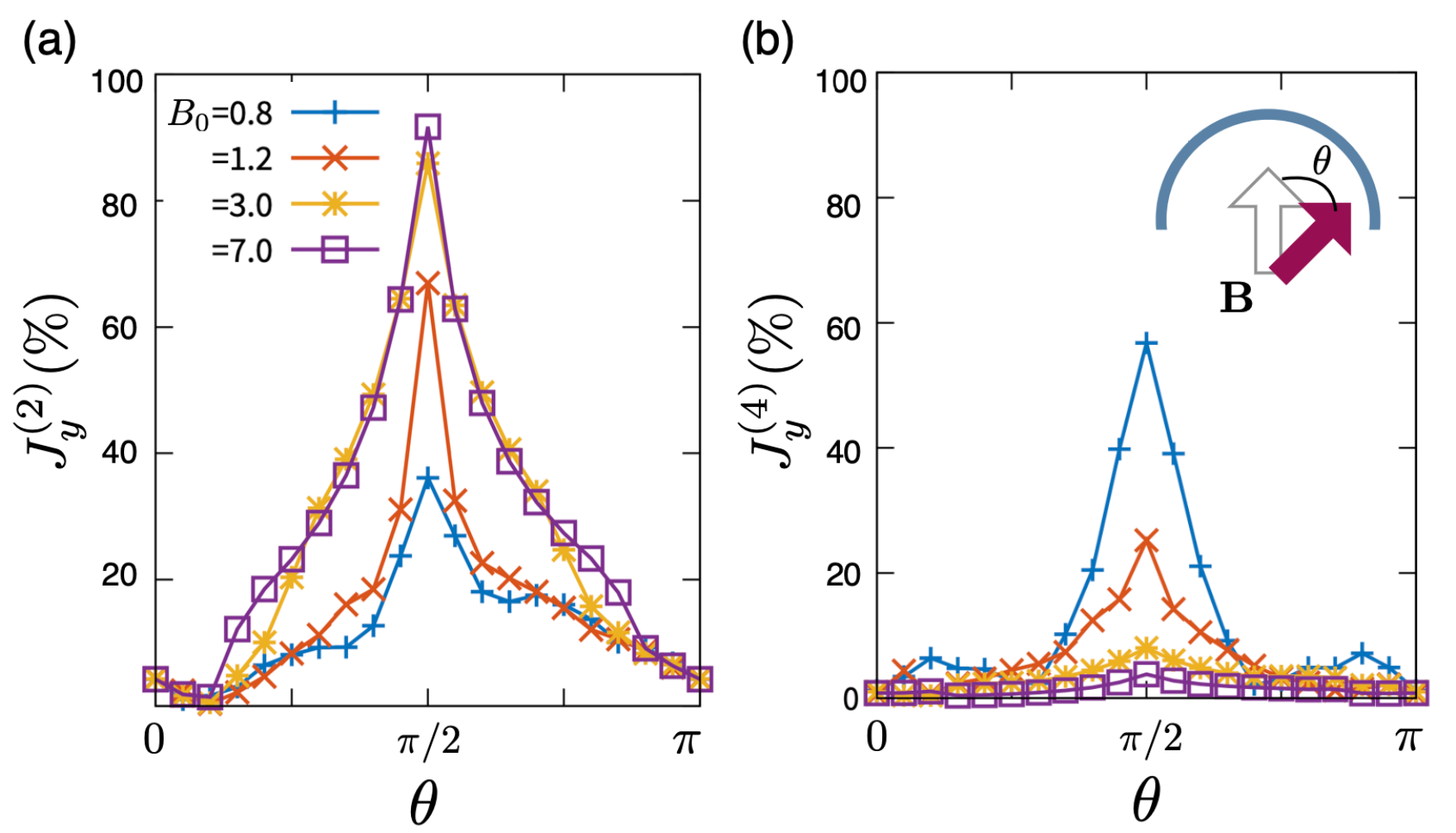}
\caption{(Color online) The relative weight of the different harmonics contributions in the Hall response as a function of the orientation of applied magnetic field ($\theta$). The blue, red, yellow and purple line presents the percentage of (a) second-harmonic Hall current and (b) fourth-harmonic Hall currents in the alternating Hall current driven by magnetic fields of different strengths $B_0 = 0.8, 1.2, 3$ and $7$ T, respectively. The alternating Hall current is estimated approximately by summing the Hall response from linear to fourth harmonic.}
\label{fig4}
\end{figure}

\begin{figure}[htp]
  \centering 
\includegraphics[width=.7\columnwidth]{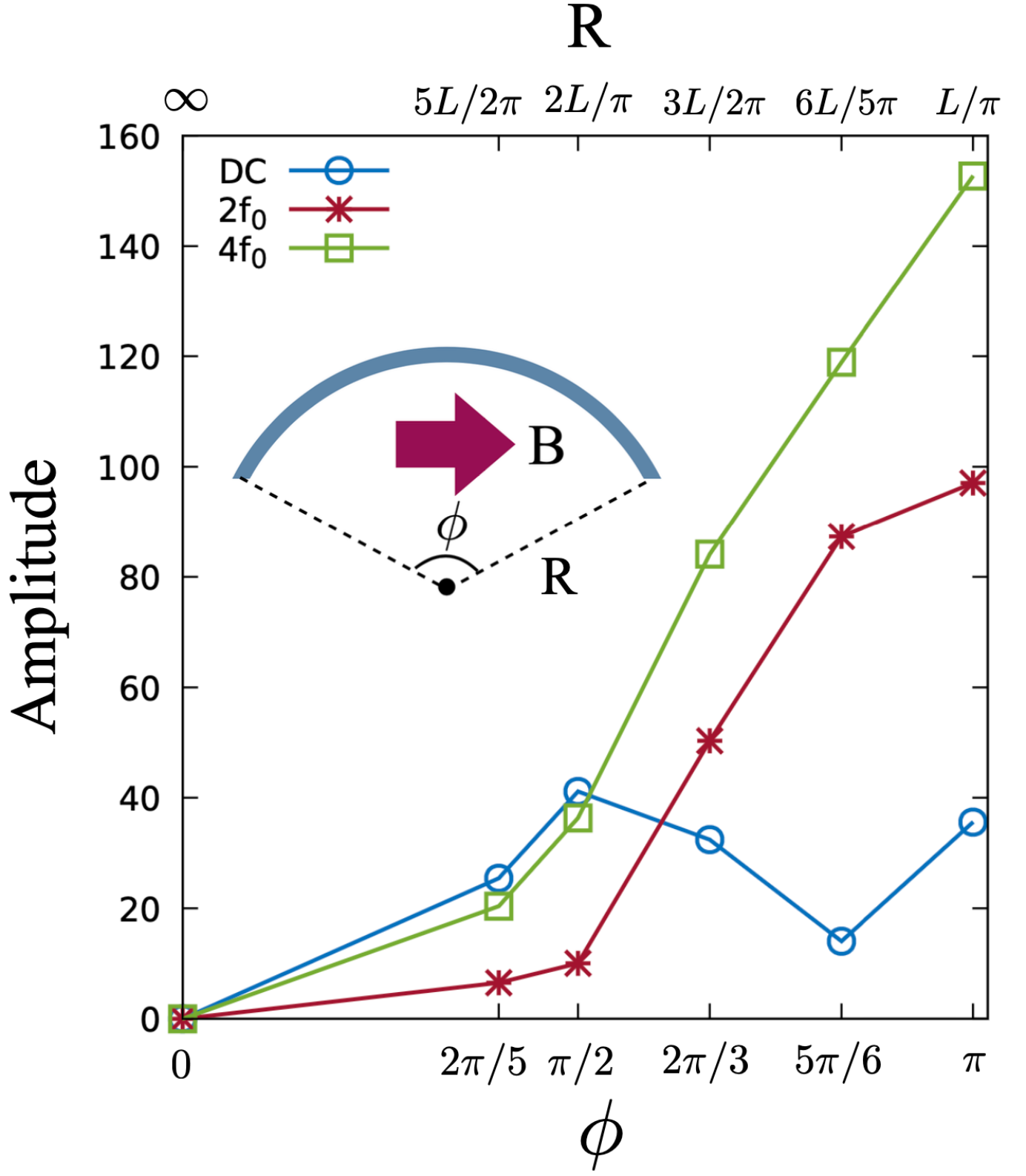}
\caption{(Color online) The amplitude of high-order Hall current decreases and increases with the radius of curvature and tangent angle, respectively. The applied field strength is $B_0=0.8$ T.}
\label{fig5}
\end{figure}

So far the focus in nonlinear Hall effect has been on second-order Hall effect which has been also experimentally observed in a variety of time-reversal-invariant material with the broken inversion symmetry \cite{Ma_2018,Kang_2019,Ho_2021,Kumar2021,Sodemann_2015,Lai2021,Gentile_2022}. Here we show the possibility of a dominant fourth-harmonic component Hall effect beyond the second-order Hall effect which is of both practical and fundamental relevance.  
On a practical level, we suggest to use curved 2D systems for on-demand generation of higher harmonics especially fourth-harmonic through magnetotransport and engineering the curvature profile. 
Moreover, the Hall effect in the curved 2D systems is anisotropic. The order of dominant Hall response can be tuned from first to fourth order by simply varying the direction and magnitude of the applied magnetic field, as is shown in Fig. \ref{fig4}.
Besides, the amplitudes of high-order harmonic components can be turned on and then increased straightforwardly by bending the curved nanomembrane to altering its curvature of radius, as is shown in Fig. \ref{fig5}.
From a fundamental viewpoint, our results provide a new avenue fundamentally to generate fourth-order Hall effect in the conducting complex nanomembrane widely, since the snake orbit leading to such effect has been manufactured in wide classes of artificial nanoarchitecture \cite{Mueller_1992, Reijniers_2000,Gentile_2022,Chang2014,Taychatanapat_2015,Rickhaus2015,Chang2017}

The prototype of the system proposed here,
can be constructed using different experimentally available settings (see Sec. S4 in the Supplemental Material). One way to have curved electronic systems is to use atomically thin 2D materials
over nonflat substrates with the desired shape and curvature \cite{reserbat2014strain,castellanos2013local,akinwande2017review}. Alternatively, one can use semiconductor heterostructures etched in a cylindrical or other curved pattern such that the resulting 2DEG layer is formed in a curved region \cite{Lorke2003,Prinz2007,vorobyova2015magnetotransport}.
In the case of 2D materials, namely graphene and graphene-like systems, there exists another interesting possibility of curvature- and strain-induced effective gauge fields which act exactly like external magnetic field with the only difference of changing sign under switching between different valleys. Apart from this difference, 
one can engineer the position dependence of 
these effective gauge fields by changing the profile of the underlying substrate and other strain-engineering techniques developed in the past decade \cite{Pereira2009,vozmediano2010gauge,si2016strain,dai2019strain}. Exploring the potential of strain-induced gauge fields in curved graphene and other 2D materials for nonlinear transport and high harmonic generation is an interesting avenue that warrants further investigation in future studies.

In a recent study, it was discovered that the use of the Berry-connection polarizability tensor enables us to go beyond the usual second-order Hall effect and obtain a third-order Hall effect \cite{Lai2021}. In this work, we demonstrate that by controlling the real-space magnetic profile through the curvature of a 2D system, we can achieve a fourth-harmonic Hall response that can be even stronger than the second-harmonic contribution. The strong fourth-harmonic signal found here indicates that the magnetic field profile due to the cylindrical curvature has a strong quadruple magnetic moment on top of its dipole. 
We further confirm that the effects occur generally in the nanostructure either with isolated or periodic corrugations. This phenomenon is entirely due to the formation of magnetic-field dipole yielding the snake orbits, a class of magnetic orbits go beyond the cyclotron orbit to drive the Hall current with higher orders. The snake states are also observed in graphene $p–n$ junctions leading to conductance oscillations \cite{Rickhaus2015,Taychatanapat_2015}, and are predicted in semiconducting core-shell nanowires subject to transversal magnetic fields \cite{rosdahl2015signature}. Therefore, our prediction can be realized not only in semiconducting curved nanostructures, but also in graphene $p–n$ junctions, core-shell semiconducting nanowires, and so on.

Finally, we would like to briefly comment on the potential expansion of the present study. Our emphasis has been on a ballistic classical transport regime, employing a semiclassical framework applicable to system sizes smaller than the mean free path but significantly larger than the Fermi wavelength. This approach furnishes a qualitative representation and substantiates the conceptual validity of the proposal. It is noteworthy that extending this investigation to diffusive and quantum transport regimes holds promise for future studies.

\section{Conclusion}\label{}
In summary, we have demonstrated that high-order Hall effects can be controlled in curved nanomembranes under transverse magnetic fields. The interplay between curvature and a constant unidirectional magnetic field generates a nonuniform effective magnetic field with a specific profile, leading to highly tuneable and versatile nonlinear Hall effects beyond the common second-order effect. Our method yields a strong fourth-harmonic Hall response due to the intricate magnetic profile that includes magnetic quadruple or higher moments in addition to magnetic dipoles. The proposed curved nanostructures for nonlinear Hall effect can be implemented in various systems, from corrugated semiconductor heterostructures to two-dimensional materials on curved substrates. Our results offer opportunities to investigate nonlinear electrical phenomena and higher-harmonic generation from both practical and fundamental perspectives.

\section*{Acknowledgements}
This work was supported in part by the Higher Education Sprout Project, Ministry of Education to the Headquarters of University Advancement at the National Cheng Kung University (NCKU). B.H., C.H.C. and J.C.Y. acknowledge the financial support by the National Science and Technology Council (Grants No. NSTC-111-2811-M-006-034-, No. NSTC-111-2112-M-006-030-, No. NSTC-112-2112-M-006-026-, and No. NSTC-112-2112-M-006-020-MY3) in Taiwan. C.H.C. would like to thank the support from the Yushan Young Scholar Program under the Ministry of Education in Taiwan.

\bibliography{ms.bib}

\end{document}


\title{Supplemental Material: \\ On-demand higher-harmonic generation through nonlinear Hall effects in curved nanomembranes}

\author{Botsz Huang}
\affiliation{Department of Physics, National Cheng Kung University, Taiwan}
\affiliation{Center for Quantum Frontiers of Research and Technology (QFort), National Cheng Kung University, Tainan 70101, Taiwan}

\author{You-Ting Huang}
\affiliation{Department of Physics, National Cheng Kung University, Taiwan}
\affiliation{Center for Quantum Frontiers of Research and Technology (QFort), National Cheng Kung University, Tainan 70101, Taiwan}

\author{Jan-Chi Yang}
\affiliation{Department of Physics, National Cheng Kung University, Taiwan}
\affiliation{Center for Quantum Frontiers of Research and Technology (QFort), National Cheng Kung University, Tainan 70101, Taiwan}

\author{Tse-Ming Chen}
\affiliation{Department of Physics, National Cheng Kung University, Taiwan}
\affiliation{Center for Quantum Frontiers of Research and Technology (QFort), National Cheng Kung University, Tainan 70101, Taiwan}

\author{\\Ali~G.~Moghaddam}\email{a.ghorbanzade@gmail.com}
\affiliation{Department of Physics, Institute for Advanced Studies in Basic Sciences (IASBS), Zanjan 45137-66731, Iran}
\affiliation{Computational Physics Laboratory, Physics Unit, Faculty of Engineering and
Natural Sciences, Tampere University, FI-33014 Tampere, Finland}

\author{Ching-Hao Chang}\email{cutygo@phys.ncku.edu.tw}
\affiliation{Department of Physics, National Cheng Kung University, Taiwan}
\affiliation{Center for Quantum Frontiers of Research and Technology (QFort), National Cheng Kung University, Tainan 70101, Taiwan}

\maketitle

\section{The methodology of simulation}

In this study, we employ the semiclassical theory within the framework of the two-dimensional electron gases (2DEGs) model. Here in our numerical model it is possible to perform that the inversion and symmetry are absent for the external field. In Eqs. (1) and (2) in the main text, ${\bf r}$ and ${\bf p}$ represent the wave-packet center of mass in real and momentum space, respectively. The subscripts are omitted since we assume a homogeneous 2DEGs model and a quadratic energy dispersion. These equations of motion adhere to the Boltzmann approach, facilitating the sampling of non-equilibrium distributions. The equations of motion offer a numerical method for sampling the distribution function of Hall current, encompassing all of its harmonic components. Importantly, this approach does not necessitate proximity to weakly nonlinear regimes or a perturbative scheme.
The numerical simulation is conducted using the Runge–Kutta methods to track the trajectories of test particles. In the post-processing stage, a discrete Fourier transform is employed to analyze the harmonic components of the Hall current. Initially, there are 2 million test particles uniformly distributed in real space, with their initial conditions determined by the Fermi circle in momentum space. The Hall current is associated with the average velocity. The methodology of simulation can be found in detail in a previous work \cite{Huang2021}.

\section{Formula for nonlinear Hall effect in strong $B$-field dipole}

In the following, the step-by-step derivation for the formula of Eq. (3) in the main article is provided. We start from the ordinary Hall conductivity in the $xy$ plane, $\sigma_{y x}\left( \pm B_0\right)= \pm e n / B_0$, which is driven by the perpendicular magnetic field $\boldsymbol{B}= \pm B_0 \hat{\boldsymbol{z}}$. The sign of field $B$ dictates the direction of the Hall current. 
When the strong magnetic-field dipole is applied in the nanomembrane, which refers to the applied negative (positive) magnetic field occurring in the left- (right-) hand side of the membrane, the net Hall current is generated approximately by the competition of two opposite Hall currents in two sides
\begin{equation}
j_y=\left(\sigma_{y x}\left(B_0\right)-\sigma_{y x}\left(-B_0\right)\right) E_x(\mathrm{t})=e\left(n_R-n_L\right) \frac{E_x(t)}{B_0}.\label{eq:1}
\end{equation}
Here $n_R$ and $n_L$ is the number of electron states per area, namely the 2D carrier density, in right and left side, respectively.

Since the carrier density proportional to the chemical potential (Fermi level), namely $n \propto \mu$, for conventional two-dimensional metallic system \cite{beenakker1991}, we directly obtain 
\begin{equation}
n_R-n_L=n \frac{\mu_R / 2-\mu_L / 2}{\mu}=n \frac{-e E_x L}{2 \mu}, \label{eq:2}
\end{equation}
where L is the width of nanomembrane. By inserting Eq. \eqref{eq:2} into Eq. \eqref{eq:1}, the Hall current is obtained as
\begin{equation}
j_y \cong-e n\left(\frac{e L E_x(t)}{2 \mu}\right) \frac{E_x(t)}{B_0}=-e^2\left(\frac{n L}{2 \mu}\right) \frac{E_x^2(t)}{B_0},\label{eq:3}
\end{equation}
which respects to the Eq. (3) in the main article.
                                                
\begin{figure*}
\centering 
\includegraphics[width=.85\textwidth]{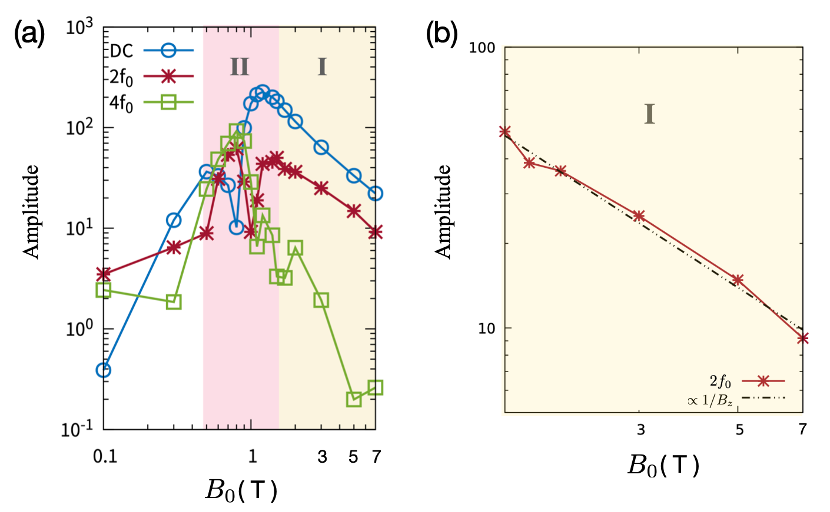}
\caption{(Color online) (a) The high-order Hall effects due to the B-field dipole, drive the current with zero, double and four times of frequencies. (b) The amplitude of current with double-times-of frequency is proportional to the inverse of the strength of magnetic-field dipole.}
\label{sfig1}
\end{figure*}

In the above equation, the magnitude of nonlinear Hall current is proportional to strength of magnetic field inversely, namely $j_y \propto 1 / B_z$, following the same role of the ordinary Hall current. Our numerical simulations also confirm this behavior, as shown in the Fig. \ref{sfig1}(b).

\section{The relation between electric field and current density for higher harmonics} 

\begin{figure*}
\centering 
\includegraphics[width=.5\textwidth]{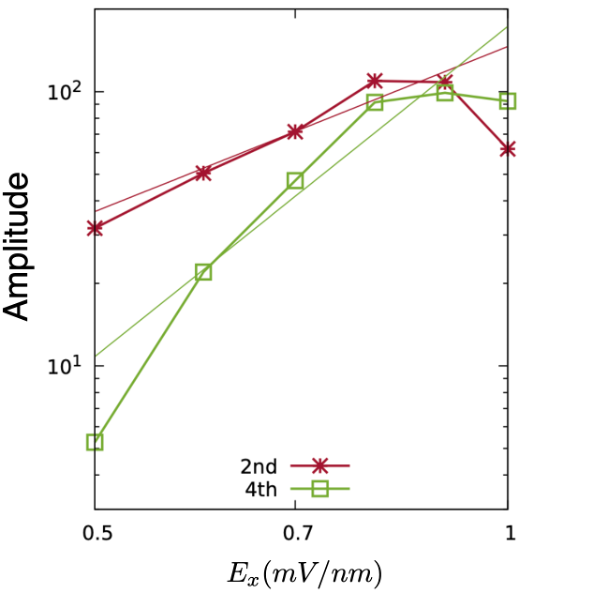}
\caption{(Color online) The amplitude of 2nd- and 4th- harmonic components of high-order Hall current driven by $B_0 = 0.8T$ in our proposed system. Both components increase with the amplitude of applied AC electric field. The red (green) dashed line is the slope proportional to $E_x^2$ ($E_x^4$), which is fitted for the 2nd- (4th-) harmonic component of Hall current. Here, $E_x$ is the amplitude of AC electric field $E_x(t)$.}
\label{sfig2}
\end{figure*}

Since the Hall current is proportional to transverse electric field, namely $j_y \propto E_x$, for ordinary Hall effect driven by the homogeneous $B$ field, the 2nd and 4th-harmonic components of nonlinear Hall current should be proportional to $E_x^2$ and $E_x^4$, respectively, when the electric field strength is moderated. Our numerical simulation of the magneto-transport of test charge particles confirms this nature of high-order components of Hall current, as is shown in Fig. \ref{sfig2}.

\section{Measurement Hall signals in suspended layered materials}

One might consider utilizing the probe arms of a cryogenic probe station or a nanomanipulation system for electrical measurements. For example, the commonly used Lakeshore probe station is capable of accommodating up to six probes, which can be effectively used to contact the four cross edges required for these types of measurements. In addition, advancements in nanofabrication now enable us to create contacts, including air bridges, for materials with suspended layers. These methods allow one to realize and measure our proposed system in practice \cite{Taychatanapat_2015,rosdahl2015signature}.

\bibliography{refs.bib}